\documentclass[aps,prb,twocolumn,groupedaddress,citeautoscript]{revtex4}

\usepackage{graphicx}
\usepackage{amsmath}
\usepackage{dcolumn}
\usepackage{colordvi}
\usepackage{times}
\graphicspath{{.}{./EPS/}}

\begin{document}

\title{Tailoring hole spin splitting and polarization in nanowires}

\author{D. Csontos}
\affiliation{Institute of Fundamental Sciences, Massey University, Private Bag
11~222, Palmerston North, New Zealand}

\author{U. Z{\"u}licke}
\affiliation{Institute of Fundamental Sciences and MacDiarmid Institute for Advanced
Materials and Nanotechnology, Massey University, Private Bag 11~222, Palmerston North,
New Zealand}
\affiliation{Institut f\"ur Theoretische Festk\"orperphysik and DFG Center for Functional
Nanostructures (CFN), Universit\"at Karlsruhe, D-76128 Karlsruhe, Germany}

\begin{abstract}
Spin splitting in \textit{p}-type semiconductor nanowires is strongly affected by the interplay
between quantum confinement and spin-orbit coupling in the valence band. The latter's
particular importance is revealed in our systematic theoretical study presented here, which
has mapped the range of spin-orbit coupling strengths realized in typical semiconductors.
Large controllable variations of the $g$-factor with associated characteristic spin polarization
are shown to exist for nanowire subband edges, which therefore turn out to be a versatile
laboratory for investigating the complex spin properties exhibited by quantum-confined holes.
\end{abstract}

\maketitle

Engineering spin splitting of charge carriers in semiconductor nanostructures may open up
intriguing possibilities for realizing spin-based electronics~\cite{sciencerev} and quantum
information processing~\cite{lossbook}. Due to the generally strong dependence of
$g$-factors on band structure~\cite{roth:pr:59}, it is expected that spatial confinement will
have an important effect on Zeeman splitting when bound-state quantization energies are
no longer negligible compared with the separation of bulk-material energy bands. The
degeneracy of heavy-hole (HH) and light-hole (LH) bulk dispersions at the zone center
makes the spin properties of valence-band states especially susceptible to such
confinement engineering~\cite{roland:prl:00,uz:prl:06,pryor:prl:06,haug:prl:06}. Recent
advances in fabrication
technology~\cite{nanoWireRev2,nanoWireRev1,defran:nlett:06,janik:apl:06,johan:natmat:06,dick:jcg:07,piccio:apl:05,romain:apl:06,oleh:apl:06}
have created opportunities to investigate hole spin physics in semiconductor nanowires
made from a range of different materials.

\begin{table}[b]
\centering \caption{\label{gammamatdep} Relative spin-orbit coupling strength $\gamma=
\gamma_{s}/\gamma_{1}$ in the valence band of common semiconductors. Here
$\gamma_{s}=(2\gamma_{2}+3\gamma_{3})/5$, and $\gamma_{1,2,3}$ denote
the Luttinger parameters~\cite{luttham2}.}
\begin{tabular}{cccccc}
\hline \hline
ZnTe/ZnS & AlAs/AlP & AlSb & CdTe & GaN/AlN & GaAs/InP \\
0.28\footnote{From Ref.~~\onlinecite{dietl:prb:01}} & 0.31\footnote{From
Ref.~~\onlinecite{vurg:jap:01}} &
0.32$^{b}$ & 0.34$^{a}$ & 0.36$^{b}$& 0.37$^{b}$\\
\hline
Ge & InN & GaSb & InAs & InSb & GaP \\
0.38$^{a}$ & 0.40$^{b}$ & 0.41$^{b}$& 0.45$^{b}$ & 0.46$^{b}$& 0.48$^{b}$ \\
\hline \hline
\end{tabular}
\end{table}
In contrast to previous theoretical
work~\cite{kossut:prb:00,kita:prb:06,xia:jpd:07,uz:prb:07b} on hole spin splitting
in quantum wires, we focus here on the influence of the spin-orbit coupling strength on
Zeeman splitting of wire-subband edges. A suitable parameter $\gamma$ quantifying
spin-orbit coupling in the valence band can be defined in terms of the effective masses
$m_{\text{HH}}$ and $m_{\text{LH}}$ associated with the HH and LH bands~\cite{warpNote},
respectively: $2 \gamma = (m_{\text{HH}} - m_{\text{LH}})/\left(m_{\text{HH}} + m_{\text{LH}}
\right)$. Table~\ref{gammamatdep} lists values for $\gamma$ in common semiconductors and
states its relation to basic band-structure parameters~\cite{luttham2}. A large part of the
theoretically possible range $0\le \gamma \le 1/2$ is covered by available materials~\cite{realNote}, 
enabling a detailed study of the interplay between spin-orbit coupling in the valence band and 
nanowire confinement. Our theoretical investigation reveals surprising qualitative differences in the 
hole spin properties of nanowires depending on the value of $\gamma$, showing that spin splitting
(and polarization) of zone-center valence-band edges in nanowires is highly tunable and has a
complex materials dependence. A detailed understanding of these properties is vital for proper
interpretation of optical and transport measurements as well as
for the design of spintronic applications involving \textit{p}-doped semiconductor nanowires.

\begin{figure}[t]
\includegraphics[width=3in]{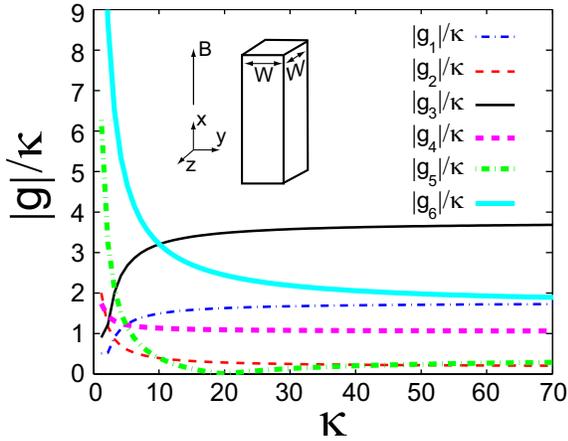}
\caption{\label{orbVSkappa}
(Color online) Effective $g$-factors for the six highest zone-center subband edges in a
GaAs wire with square cross-section, plotted as a function of the bulk-hole $g$-factor
$\kappa$. An order of magnitude enhancement in $\kappa$ leads to saturation, in effect
quenching orbital contributions to the Zeeman splitting.}
\end{figure}

We use the Luttinger model~\cite{luttham2} in the spherical approximation~\cite{lip:prl:70} for the
top-most bulk valence bands. Including the bulk Zeeman term $H_{\text{Z}} =-2
\kappa \mu_{\text{B}} B \hat{J}_{z}$, the Hamiltonian is given by
\begin{equation}
H=-\frac{\gamma_{1}}{2m_{0}}p^{2}+\frac{\gamma_{s}}{m_{0}}\left [ ({\mathbf p}\cdot \hat
{\mathbf J})^{2}-\frac{5}{4}p^{2}{\mathbf 1}_{4\times 4} \right ]+H_{\text{Z}} \quad .
\label{Hamiltonian}
\end{equation}
Here ${\mathbf p}$ is the linear orbital momentum, $\hat{\mathbf J}$ the vector of spin-3/2
matrices, $m_{0}$ the electron mass in vacuum, $\gamma_{s}=(2\gamma_{2}+3 \gamma_{3})
/5$ in terms of the Luttinger parameters~\cite{luttham2}, $\mu_{\text{B}}$ is the Bohr
magneton and $\kappa$ the bulk hole $g$-factor. We neglect the small anisotropic part of the
bulk-hole Zeeman splitting. A hard-wall confinement in the $xy$ plane defines the quantum
wire with either cylindrical or square cross-section. Our method for finding the zone-center
subband edges and calculating their $g$-factor $g^\ast$ in a magnetic field parallel to the wire
axis has been described elsewhere~\cite{uz:prb:07b,uz:physe:07b}. An intriguing universal
behavior of wire-subband spin splittings emerges when the bulk-Zeeman term dominates the
orbital effects which, in principle, also contribute to the effective $g$-factor. This universal regime,
which is characterized by $g^\ast$ scaling with $\kappa$ and being independent of wire diameter,
is accessible in real nanowire systems~\cite{defran:nlett:06} where $\kappa$ is enhanced by the
\textit{p-d} exchange interaction with magnetic acceptor ions~\cite{dietl:prb:01}.
Figure~\ref{orbVSkappa} illustrates that, for the highest (i.e., closest to the top of the valence band)
GaAs hole-wire levels, only a moderate enhancement of $\kappa$ is needed to quench orbital
contributions to the $g$-factor. Similar results are obtained for other materials. In the following, we
focus entirely on the properties of hole-wire subband-edge $g$-factors in the universal regime
where orbital contributions can be neglected.

\begin{figure}[b]
\includegraphics[width=3in]{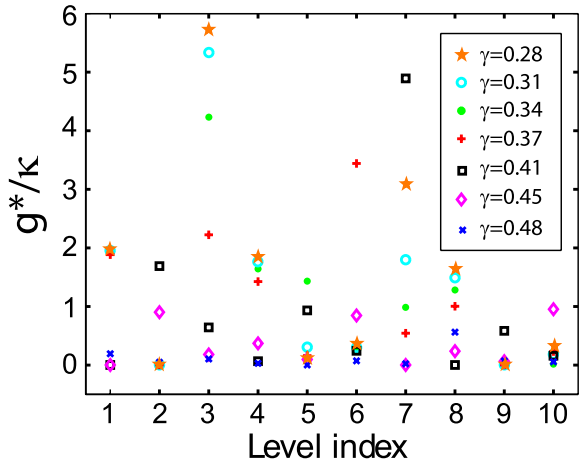}
\caption{\label{gFactResults}
(Color online) Effective $g$-factors for the ten highest zone-center subband edges in
cylindrical hole nanowires, calculated for various spin-orbit coupling strengths.}
\end{figure}
Our results are summarized in Figure~\ref{gFactResults} where we show $g$-factors
for the ten highest zone-center subband edges in
cylindrical hole nanowires, calculated for various spin-orbit coupling strengths $\gamma$.
A na\"{\i}ve assumption that the hole spin projection parallel to the wire axis should be
quantized would lead us to expect to find only two possible values for the $g$-factor;
namely $6\kappa$ and $2\kappa$ for the HH and LH states, respectively. Evidently, our
results are quite different. Firstly, for any given material, the $g$-factor values vary strongly
between the different wire-subband edges, some levels even displaying vanishing
$g$-factors. Such seemingly random fluctuations can be
explained~\cite{uz:prb:07b,uz:physe:07b} by nontrivial microscopic hole spin-polarization
profiles of wire-subband bound states. Large $g$-factors are found for subband edges with
predominantly HH or LH character, whereas subbands with mixed HH-LH character or with
vanishing hole-spin polarization have strongly suppressed $g$-factors. We will see below
that the intrinsic connection between hole spin splittings and polarizations holds for all
materials considered. Secondly, focusing on individual wire levels, it is found that their
$g$-factor can vary substantially between different materials. For some subbands, e.g.,
the third and seventh, the $g$-factors span almost the entire range of values between $0$ and $6
\kappa$. For other subbands, $g$-factors cluster around certain values, as is the case of
the first, sixth, and tenth levels. Yet other subbands display a seemingly random sequence
of alternatingly increasing and decreasing values of $g^{\ast}$ as the relative spin-orbit
coupling strength $\gamma$ is varied.

\begin{figure}[b]
\includegraphics[width=3in]{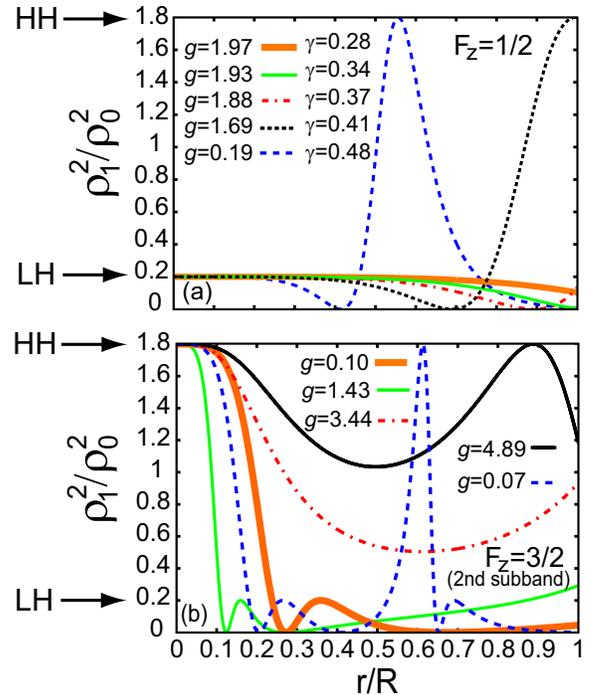}
\caption{\label{SpinPolProfile}
(Color online) Squared normalized spin-3/2 dipole (spin-polarization) density, $\rho_{1}^{2}(r)/
\rho_{0}^{2}(r)$, for (a) the highest subband with $F_z=1/2$, and (b) the second-highest subband
with $F_z=3/2$. The values of spin-orbit coupling parameter $\gamma$ and corresponding
$g$-factor $g\equiv g^\ast/\kappa$ are indicated.}
\end{figure}
The anomalous spin splittings in hole nanowires can be attributed to strong HH-LH
mixing that is present even at the wire-subband edges. The relative spin-orbit coupling strength
$\gamma$ determines this mixing. To be able to characterize the spin properties of individual
subband-edge bound states independent of any particular spin-projection basis, we utilize
scalar invariants of the spin-3/2 density matrix. See Refs.~~\onlinecite{roland:prb:04,uz:prb:07b}
for details of the formalism. In particular, we consider the radial variation of the normalized
hole-spin dipole density, denoted by $\rho_{1}^{2}/\rho_{0}^{2}$, which provides a measure of
the local hole spin polarization.  A uniform value of $\rho_{1}^{2}/\rho_{0}^{2}=9/5$ ($1/5$)
indicates a HH (LH) state characterized by a ${\hat J}_z$-projection quantum number $\pm 3/2$
($\pm 1/2$). As previously discussed, Zeeman splitting for such a state in a magnetic field
parallel to the $z$ axis arises with effective $g$-factor $6\kappa$ ($2\kappa$)~\cite{luttham2}.
Figure~\ref{SpinPolProfile} shows the radial spin-polarization profiles $\rho_{1}^{2}(r)/\rho_{0}^{2}
(r)$, for the highest hole-wire subband edges with (a) $F_z=1/2$, and (b) the second-highest
subbands with $F_z=3/2$, for different representative values of $0.28\leq \gamma \leq 0.48$.
Here, $F_z$ is the eigenvalue of $\hat J_z + \hat L_z$, i.e., the sum of the $z$ components of
spin and orbital angular momentum, which is the good quantum number labelling wire-subband
bound states~\cite{sercel:apl:90,uz:prb:07b}. Deviations of the hole-spin polarization from the
values $9/5$ and $1/5$ is an indication of the, in principle, ever-present HH-LH mixing in hole
wires.

Interestingly, states with $F_z=1/2$ that form the highest subband edge in systems with
$\gamma\le 0.37$ are quite close to a pure LH character, having $\rho_{1}^{2}(r)/\rho_{0}^{2}(r)
\approx 0.2$ across most of the wire radius. However, a continuously increasing trend to develop
a HH-LH texture is exhibited for larger $\gamma$. As can be seen, this feature is concomitant
with a drastic reduction of the $g$-factor from its value close to $2\kappa$ that is expected for
pure LH states. A related trend is exhibited by the highest subband edges with $F_z=3/2$ (not
shown here) where, for small values of $\gamma$, the normalized dipole moment is close to the
value $9/5$ corresponding to a pure HH state. With increasing $\gamma$, however, the dipole
moment is increasingly suppressed. The $g$-factors show a corresponding monotonous
suppression, from values close to $6\kappa$ to values close to 0.

\begin{figure}[t]
\includegraphics[width=3in]{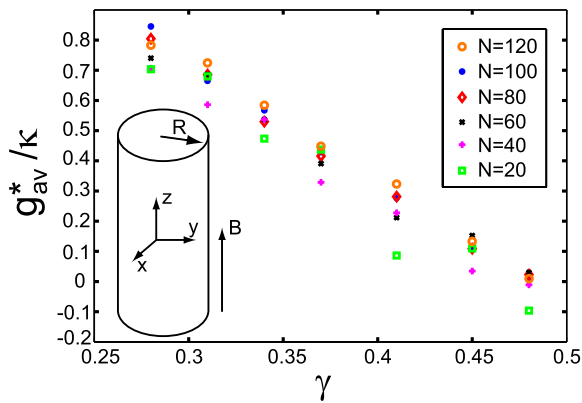}
\caption{\label{avGfacts}
(Color online) Mean g-factors $g_{\text{av}}^{\ast}=\frac{1}{N}\sum_{i=1}^{N} g^\ast_{i}$,
obtained by averaging over the $N$ highest wire levels, plotted as a function of relative
spin-orbit coupling strength $\gamma$. Inset: Wire geometry and orientation of the magnetic
field.}
\end{figure}
In contrast to the previous two examples, a very nonmonotonous behavior as a function of
$\gamma$ is observed for the second-highest subband edge with $F_z=3/2$. See
Fig.~\ref{SpinPolProfile}(b) where, for small $\gamma$-values, suppressed polarization
profiles correlate with very small effective $g$-factors. As $\gamma$ is increased, the spin dipole
moment of the state increases dramatically, approaching values associated with HH character. 
[See the dashed-dotted and dashed curves corresponding to $\gamma=0.37,0.41$ in
Fig.~\ref{SpinPolProfile}(b).] The corresponding $g^\ast$ values come close to $6\kappa$.
For yet higher values of $\gamma$, the polarization is again suppressed, with concomitantly
vanishing $g$-factors.

A general comparison of polarization profiles for various subband edges with their $g$-factors
shows that, as the hole-spin dipole moment vanishes and/or HH-LH mixing in the radial profile
increases, $g^\ast$ is increasingly suppressed. Thus, a direct correlation emerges between the
relative spin-orbit coupling strength $\gamma$, the hole-spin polarization, and the Zeeman spin
splitting. However, on average, the hole-spin polarization and effective $g$-factors decrease as
the relative spin-orbit coupling strength $\gamma$ is increased. This is illustrated by the calculated
mean $g$-factors shown in Fig.~\ref{avGfacts}. Such mean values will describe Zeeman splitting in
experimental situations where single wire subbands are not resolved. Extrapolating to
$\gamma=0.38$, which corresponds to Ge, the value found is consistent with the hole $g$-factor
measured recently~\cite{stefanoSUB} in rod-shaped quantum dots fabricated from Ge/Si core-shell
nanowires.

DC acknowledges support from the Massey University Research Fund. The authors benefited from
useful discussions with P.~Brusheim, A.~F\"uhrer, S.~Roddaro, and H.Q.~Xu.


\begin{thebibliography}{25}
\bibitem{realNote}
In the following, we present results for $0.28 \le \gamma \le 0.48$ because nanowires have
recently been grown from ZnTe~\cite{janik:apl:06} ($\gamma=0.28$), GaAs~\cite{defran:nlett:06}
($\gamma=0.37$), Ge~\cite{stefanoSUB} ($\gamma=0.38$), InAs~\cite{dick:jcg:07}
($\gamma=0.45$), and GaP~\cite{johan:natmat:06} ($\gamma=0.48$).
\end{thebibliography}

\end{document}